\documentclass[aps,epsf,showpacs,preprint]{revtex4}
\usepackage[latin1]{inputenc}
\usepackage{graphics}
\usepackage{amsfonts, amssymb, latexsym, amsmath}
\input epsf.sty
\begin{document}
\title{Force-induced desorption of a linear polymer chain adsorbed on an attractive surface}
\author{P. K. Mishra, Sanjay Kumar and Yashwant Singh}
\email{pramod@justice.com, yashankit@yahoo.com, ysingh@bhu.ac.in}
\affiliation{Department of Physics \\
Banaras Hindu University \\
Varanasi-221 005, India}
\begin{abstract}
We consider a model of self-avoiding walk on a lattice with on-site repulsion and an attraction
for every vertex of the walk visited on the surface to study force-induced desorption of a
linear polymer chain adsorbed on an attractive surface and use the exact enumeration technique
for analyzing how the critical force for desorption $f_c(T)$ depends on the temperature.
The curve $f_c(T)$ gives the boundary separating the adsorbed phase from the desorbed phase.
Our results show that in two dimensions where surface is a line the force $f_c(T)$ increases
monotonically as temperature is lowered and becomes almost constant at very low temperatures.
In case of three-dimensions we, however, find re-entrance, i. e. $f_c(T)$ goes through
a maximum as temperature is lowered. 
The behaviour of the polymer chain at different values of temperature and force is examined by 
calculating the probability distribution of the height from the surface of the vertex at which
external force is applied.
\end{abstract}

\pacs{68.18.Jk,68.47.Pe,36.20.-r}

\maketitle

When a long flexible polymer chain interacts with an impenetrable surface its conformational
properties are strongly modified in comparison with its bulk properties \cite{Eisen,tri}.
This is due to a subtle competition between the lowering of internal energy near an attractive
surface and the loss of entropy due to constraints imposed by the impenetrable surface. For
a strongly attractive surface, the polymer chain sticks to the surface, and for weak attraction
it prefers to stay away from the surface. Thus there is a transition from the state when chain
is mostly attached to the surface (adsorbed) to the state of detachment (desorbed) when the
temperature is increased. The transition between theses two states is marked by a transition 
temperature $T_a$ with the adsorbed phase for $T<T_a$ and the desorbed phase for $T>T_a$. 

A model of self avoiding-walk on a lattice with on-site repulsion and attraction energy
for every vertex of the walk visited on the surface provides an adequate model for 
understanding the adsorption-desorption transition \cite{YSingh,YSingh1,YSingh2}. 
We extend this model of 
self-avoiding walk to study the force-induced desorption of a linear polymer chain adsorbed
on an attractive surface and calculate the critical force $f_c(T)$ for desorption as a 
function of temperature. Response of a polymeric chain to externally applied force can be 
measured experimentally by using techniques like optical or magnetic tweezers \cite{opt} 
and atomic force microscope \cite{afm}.

We consider self-avoiding walks $(SAWs)$ that start from a point on an impenetrable surface and
experience a force $f$ in a direction perpendicular to the surface at the other end. The 
applied force, because of its direction, favours desorption and one expects a critical
force, $f_c(T)$, for desorption.
At a given $T$ when the applied force $f$ is less than $f_c(T)$ the polymer will
be adsorbed while for $f>f_c(T)$ the polymer will be desorbed. The curve $f_c(T)$, therefore,
gives the boundary that separates the desorbed phase from the adsorbed phase in the $(f,T)$
plane.

Let $z=0$ represents the surface and walks start from a point (origin) on the surface. 
In case of two dimensions (2-$d$) the surface is a line whereas in case of three dimensions
(3-$d$) the surface is a plane. Since the surface is impenetrable, walks are restricted to
half of the space $(z\ge 0)$ only. We enumerated all $SAWs$ upto a
certain length on a square lattice in 2-$d$ and on a cubic lattice in 3-$d$.
Let $C_N(N_s,h)$ be the number of $SAWs$ of $N$ vertex (a vertex of the walk represents a 
monomer of the polymer and a step of the walk, the chemical bond connecting the 
neighbouring monomers)
having $N_s$ number of vertices on the surface, $N-N_s$ vertices away from the surface and
$h$ the height of the end vertex of the walk from the surface. 
We enumerated and analyzed the series $C_N(N_s,h)$ upto $N\le 31$ in 2-$d$ and $N\le 20$
in 3-$d$. The values of $C_N(N_s,h)$ found from the datas of exact enumerations for a given $N$
with all possible values of $N_s$ and $h$ are given in 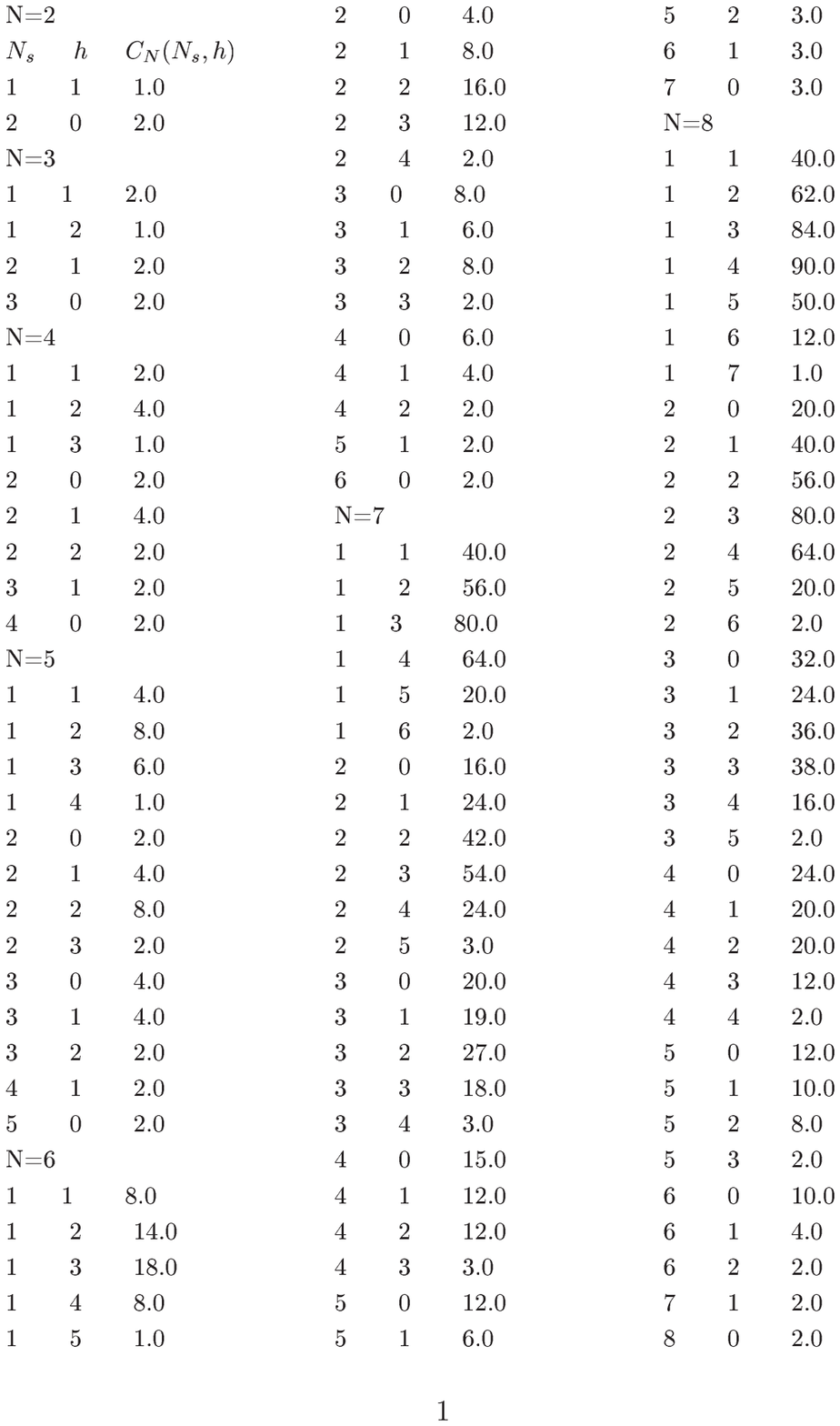 and 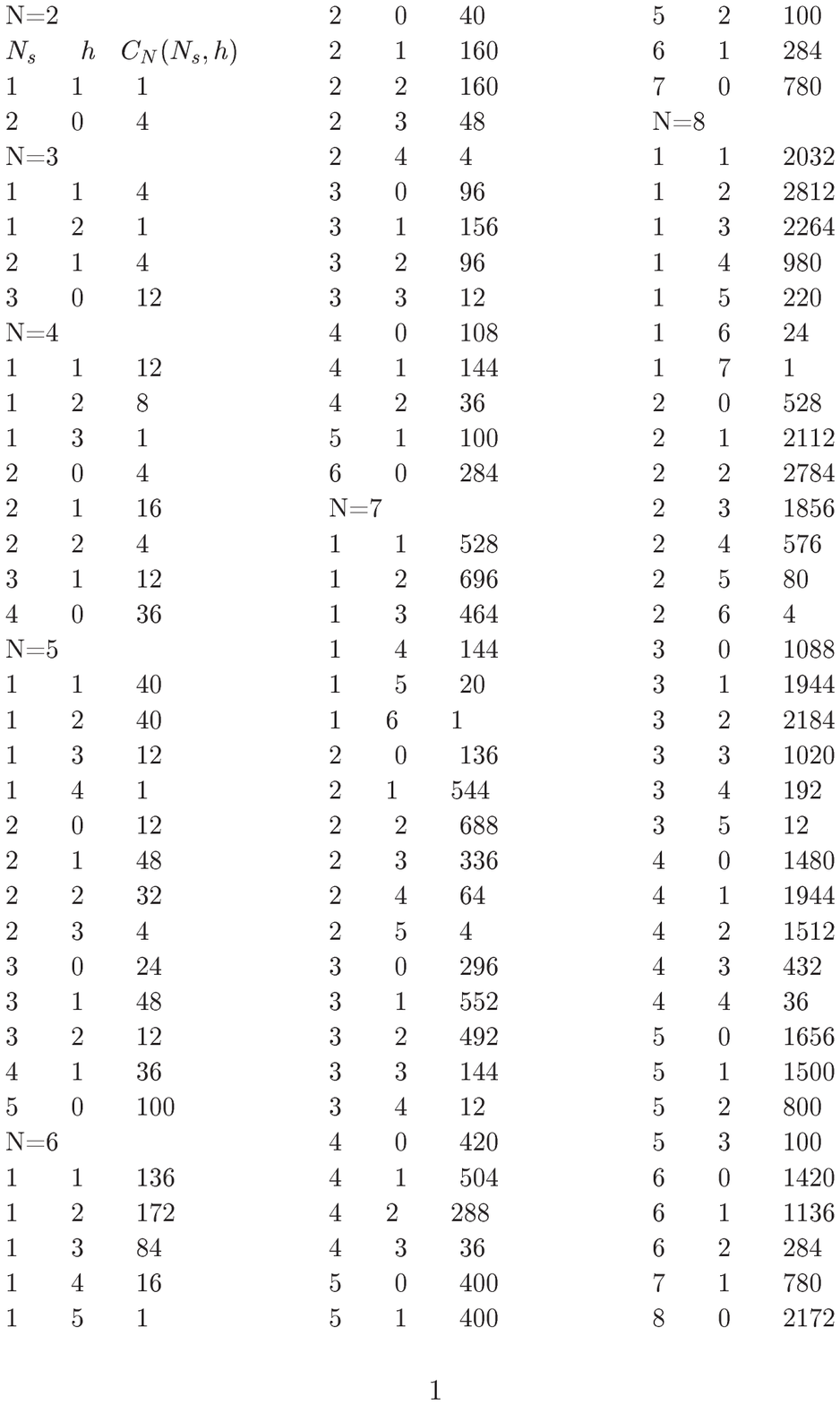 (see source
file draft.tgz). 

The partition function is found from the relation 
\begin{equation}
\label{e.1}
Z_N(\omega,u)=\sum_{N_s,h} C_N(N_s,h) {\omega}^{N_s}u^h
\end{equation}
where $\omega=e^{-\epsilon_s/k_B T}$ and $u=e^{f/k_B T}$. Here
$\epsilon_s$ is the energy of attraction ($\epsilon_s <0$) of a vertex (or monomer) with the 
surface and $f$ is the force acting at the last vertex of the walk in a direction perpendicular
to the surface. In what follows, we set the Boltzmann constant $k_B =1$ and $\epsilon_s =-1$ and 
express the length in unit of the lattice parameter or the length of a step of the walk.

For fixed $f$ we locate the adsorption-desorption transition temperature from the maximum 
of $\frac {\partial <N_s>}{\partial (\log \omega)}(=<{N_s}^2>-<N_s>^2)$ where 
$<N_S>=\frac{1}{Z_N(\omega,u)} \sum_{N_s,h} N_sC_N(N_s,h) {\omega}^{N_s}u^h$ (see Eq. (8)).  In Fig. (1) we 
plot the values of $\frac {\partial <N_s>}{\partial (\log \omega)}$
as a function of $T(=\frac {1}{\log \omega})$ for some values of $f$. The results 
given in this figure correspond to walks ($a$) of 31 vertices in 2-$d$ and $(b)$ 
of 20 vertices in 3-$d$. 
The curve $\frac {\partial <N_s>}{\partial (\log \omega)}$ vs $T$ of 2-$d$ is found to
have  only one
maximum for all values of $f$ for which the adsorption-desorption transition occurs on varying
the temperature, but in 3-$d$ the curve for certain values of $f (\ge 1)$ is found to have 
two maxima.
The occurrence of two maxima in a $\frac {\partial <N_s>}{\partial (\log \omega)}$ curve
indicates that the adsorption-desorption transition takes place at two different temperatures
for the same value of $f$. In such a case the adsorbed phase is bounded by the 
desorbed phase from both low and high temperature sides. This is the case of a re-entrance.

It is possible to obtain better estimates of phase boundaries by extrapolating for
large $N$. The reduced free energy per monomer defined as
\begin{equation}
\label{e.1}
G(\omega,u)= \lim_{ N\to \infty} \frac{1}{N} \log Z_N(\omega,u)
\end{equation}
can be estimated from the partition functions found from the data of exact enumerations for finite $N$ by extrapolating
to large $N$. For $N\to \infty$ one can, in general, write \cite{tri},
\begin{equation}
\label{e.4}
Z_N(\omega,u)\sim {\mu(\omega,u)}^NN^{\gamma -1}
\end{equation}
where $\mu(\omega,u)$ is the effective coordination number and $\gamma$ is the universal
configurational exponents for walks with one end attached to the surface. The value of 
$\mu(\omega,u)$ can be estimated using ratio method \cite{ratio} with associated 
Neville table or any other method such as Pad\`e analysis \cite{pade} or differential 
approximants
\cite{da}. The ratio method has recently been used to predict the
phase diagram of a long flexible polymer chain immersed in a poor solvent near an attracting
surface \cite{YSingh1,YSingh2}.

From Eqs. (2) and (3) we can write 
\begin{equation}
\label{e.3}
\log \mu(\omega,u)=\lim_{ N\to \infty} \frac{1}{N} \log Z_N(\omega,u)=G(\omega,u)
\end{equation}
$Z_N(\omega,u)$ is calculated from the data of $C_N(N_s,h)$ using Eq. (1) for given
values of $\omega$ and $u$ ( or equivalently, $T$ and $f$). From this we construct linear and
quadratic extrapolants of the ratio of $Z_N(\omega,u)$ for adjacent values of $N$ as
well as the alternate one. The values of $\mu(\omega,u)$ thus obtained for some values of
force and temperature are shown in tables $I$ and $II$ for 2-$d$ case and in table $III$ and
$IV$ for 3-$d$ case. As shown in Figs (2) and (3) and from the tables(I-IV), the results for 
alternate $N$ give better convergence. 
We used this value of $G(\omega,u)$ to locate the transition temperature for given $f$
from the maxima of $\frac{{\partial}^2 G}{\partial {(\log\omega)}^2}$(=
$\frac {\partial <N_s>}{\partial (\log \omega)}$). 

The phase diagram thus obtained is shown
in Fig. (4). We also plot in this figure the results found from the partition function
of a chain of finite length; $N=31$ in 2-$d$ and $N=20$ in 3-$d$. Except for small difference 
seen at high temperatures the agreement between these two values is good.
We note that in case of 2-$d$  
the critical force for desorption increases monotonously as temperature is lowered and becomes
constant at very low temperatures, $T\le 0.29$. In case of 3-$d$ we find, as mentioned
above, re-entrance i. e. 
the critical 
force goes through a maximum as $T$ is lowered and the adsorbed phase for $f\ge 1$ is bounded from both low and high temperature sides by the desorbed phase.
The maximum of $f_c(T)$ occurs at $T\sim 0.77$ and its value is found to be 1.5, 
(see Fig. $4(b)$). In both 2-$d$ and 3-$d$ cases the value of  $f_c$ at $T=0$ is 1.
The re-entrance in phase diagram has also been reported in 3-$d$
in directed walk models \cite{Somen,Marenduzzo,Marenduzzo1}.

The nature of the phase boundary at low temperatures as shown in Fig. 4 is perhaps due to
the specific configuration acquired by the part of chain that gets detached from the 
surface under the influence of the external force. As shown below, at low temperatures 
when the applied force is close to its critical value
the detached part of the chain
becomes a rod oriented along a particular direction. There is, therefore, no entropic 
contribution to the free energy from this part of the chain. On the other hand, at high 
temperatures the part of the chain that gets detached under the influence of the external 
force remains as a self-avoiding walk in the half space with all lengths thermal 
fluctuations unless the force becomes sufficiently large compared to the critical force.  
 
To see this we
calculate the probability distribution of $h$ defined as
\begin{equation}
\label{e.3}
P(h)=\frac{u^h}{Z_N(\omega,u)}\sum_{N_s}C(N_s,h) {\omega}^{N_s}
\end{equation} 
for different values of $f$ and $T$. The results are shown in Fig. 5 for 2-$d$ and in 
Fig. 6 for 3-$d$ at two temperatures. 

These figures show the difference in the behaviour of a polymer in the low and high temperature
regions of the phase diagram. At high temperatures the thermal fluctuations
are present in both the adsorbed and desorbed parts of the chain. As a consequence, even 
at $f=0$ when the chain is fully adsorbed, $P(h)$ has a large width and its value at $h=0$ is
less than $0.5$. It is only when $f$ is sufficiently large than $f_c$, the polymer chain gets
stretched under the influence of force, as the force suppresses the thermal 
fluctuations. On the other hand, at low temperatures ($T\sim 0.1$) in both, 2-$d$ and 3-$d$,
the polymer chain seems to fully lie on the surface with $h\sim 0$ at $f<f_c$. It is only when 
$f$ is close to $f_c$, $P(h)$ gets broadened indicating that a part of the chain is 
pulled away from the surface. For $f>f_c$ the polymer chain seems to acquire   
shape of a rod in the direction perpendicular to the surface.
At $f\sim f_c$ all values of $h$ appears equally probable indicating large fluctuations in the
segment of the polymer that lies on the surface and the other segment that is away from the 
surface under the influence of the force.

We calculated the entropy per monomer for $f=0$ using the relation 
\begin{equation}
\label{e.4}
s=\frac{\partial (TG(\omega,u=1))}{\partial T}
\end{equation}
and found (see Fig. ($7$)) that in 2-$d$ the entropy goes to zero for $T\le 0.27$ 
indicating that the chain is fully adsorbed and acquires a shape of one dimensional object.
This may, however, not be strictly true in a long chain $(N\to \infty)$ due to presence 
of low energy long wavelength thermal fluctuations which are suppressed in a short chain.
In 3-$d$ entropy becomes constant equal to $s_a= \ln\mu=0.97$, ($\mu$ being the 
connectivity of a square lattice) for $T\le 0.40$. This shows that for $T\le 0.4$ 
the polymer chain is
fully adsorbed on the surface and acquires a configuration of a chain in two-dimensions.
It therefore seems that below certain temperature the polymer chain is fully adsorbed and
confined to the surface.

From the above results one concludes that when the value of the external force $f$ is close to 
the critical value $f_c$ a surface adsorbed polymer chain at very low temperatures ($T\sim 0$)
has two parts; one is fully adsorbed on the surface or
zipped with the surface and the other is in a shape of rod perpendicular to the surface.
Thus near $T\to 0$ we can write the free energy as \cite{Marenduzzo,Marenduzzo1}
\begin{equation}
\label{e.5}
F= -N_s -T N_s s_a -f (N-N_s)
\end{equation} 
where $N-N_s$ represents the length $h$ perpendicular to the surface; $s_a$ is the
entropy associated with the adsorbed state.
Minimization of Eq. (7) with respect to $N_s$ gives $f_c(T)= 1+T s_a$ and $\frac{df_c}{dT}= s_a$.
The value of $\frac{df_c}{dT}$ at $T\to 0$ is found from Fig. $4(b)$ to be $\sim 0.93$ 
which is in good agreement with the value $s_a=0.97$.
Since in case of 2-$d$, $s_a$ is zero as the surface is a line we get $f_c=1$ and 
$\frac{df_c}{dT}=0$ in agreement with the result shown in Fig. $4(a)$.

The values of $\frac{<N_s>}{N}$ and $\frac{<h>}{N}$ calculated from relations 
\begin{equation}
\label{e.8}
\frac{<N_s>}{N}=\frac{1}{Z_N(\omega,u)}\frac{1}{N} \sum_{N_s,h} N_s C_N(N_s,h){\omega}^{N_s}u^h
\end{equation}
and
\begin{equation}
\label{e.9}
\frac{<h>}{N}=\frac{1}{Z_N(\omega,u)}\frac{1}{N} \sum_{N_s,h} h C_N(N_s,h){\omega}^{N_s}u^h
\end{equation}
and plotted in Fig. 8 show a sharp jump; a characteristic of a first-order
transition. The result that the force-induced adsorption-desorption is first-order
has also been found in the directed walk model 
by Orlandini et al \cite{Somen}. 

In this paper we have used the lattice model of self-avoiding walk to calculate the values
of critical force $f_c(T)$ for desorption of a linear polymer chain adsorbed on an attractive
surface at different temperatures. The phase boundary that separates the adsorbed phase
from the desorbed phase shows re-entrance in case of 3-$d$. It is found that for $f\ge 1$
the adsorbed phase is bounded from both the low and high temperature sides by the desorbed
phase. In case of 2-$d$ we, however, find that the critical force $f_c(T)$ attains a constant
value for $T\le 0.29$ and does not change on lowering the temperature. These features of the
phase diagram can be understood from the configurations the part of the polymer chain that
gets detached from the surface under the influence of the force acquires at the low 
temperatures. We have calculated the probability distribution of the height from the surface
of the end vertex at which external force is applied at different values of 
force and temperature
and found that for $T\sim 0$ and $f\sim f_c$ the polymer chain has two parts;
one is fully adsorbed on the surface and the other is in a shape of a rod perpendicular
to the surface.

\acknowledgments
Financial support from $CSIR$, $INSA$ and $DST$ New Delhi, India thankfully acknowledge.

\newpage

\begin{figure}[htp!]
\begin{center}
\setlength{\unitlength}{1cm}
\begin{picture}(5,8)
\put(-1,-.1){\scalebox{.8}[.8]{\hspace{-5cm}\includegraphics*{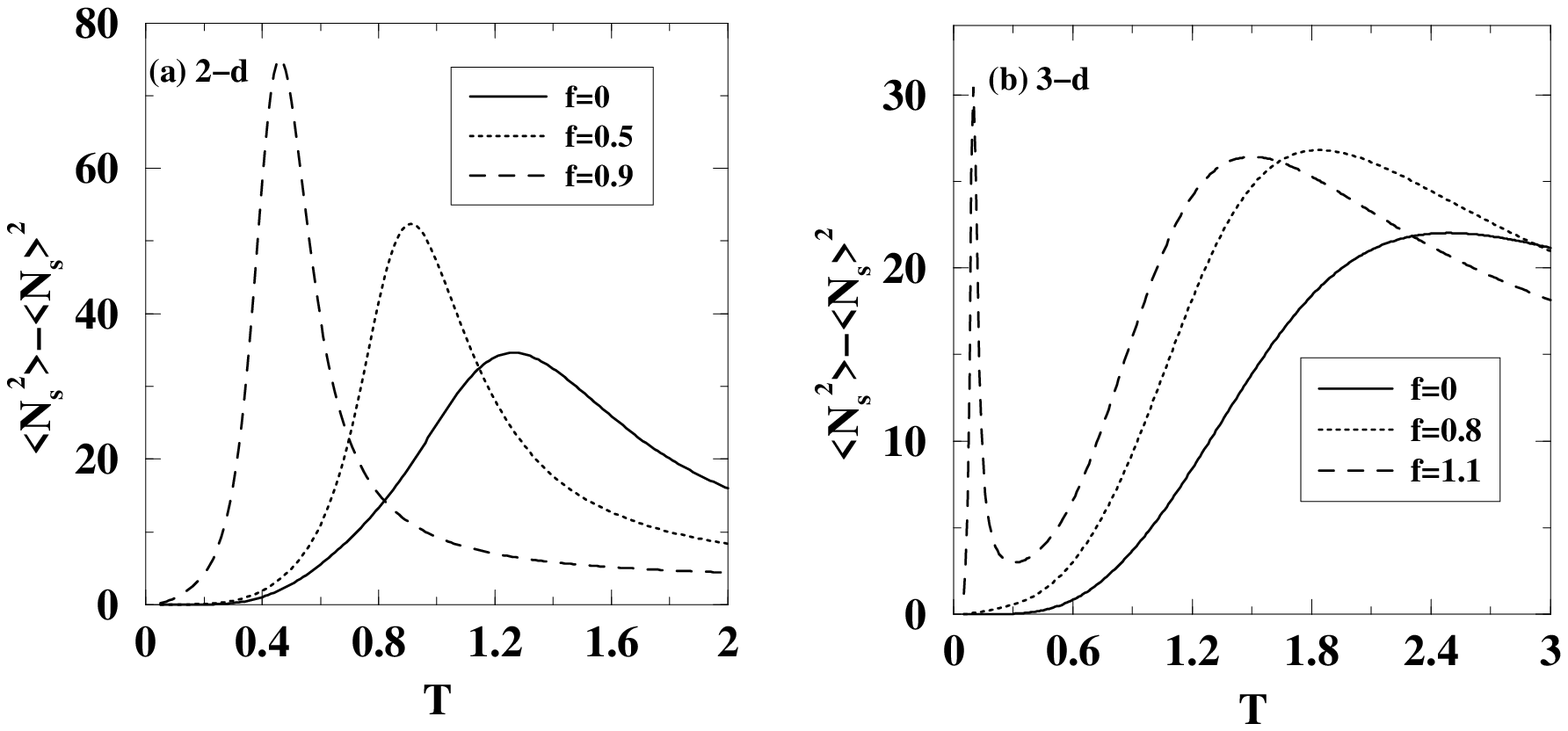}}}
\end{picture}
\caption[a]{ The dependence of 
$\frac {\partial <N_s>}{\partial (\log \omega)}= <{N_s}^2>-<N_s>^2$ on $T$ for 
different forces is shown. Results are $(a)$ for $N=31$ in 2-$d$ and $(b)$ for
$N=20$ in 3-$d$. Note the existence of two peaks for $f=1.1$ in 3-$d$.}
\label{1f}
\end{center}
\end{figure}
\vspace{1.2cm}

\begin{figure}[htp!]
\begin{center}
\setlength{\unitlength}{1cm}
\begin{picture}(5,6)
\put(-1,-.1){\scalebox{.8}[.8]{\hspace{-5cm}\includegraphics*{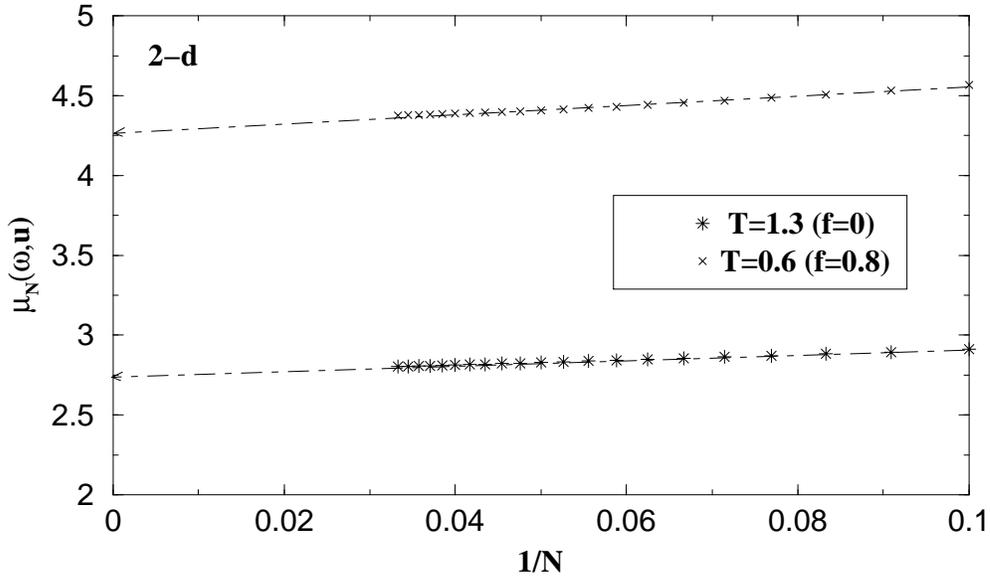}}}
\end{picture}
\caption[a]{ The value of $\mu_N(\omega,u)$ calculated from the relation 
$\mu_N(\omega,u) =\sqrt{\frac{Z_N(\omega,u)}{Z_{N-2}(\omega,u)}}$ in 2-$d$ 
is plotted as a function 
of $\frac{1}{N}$ at values of ($f,T$)=(0,1.3) and (0.8,0.6). The extrapolated value
of $\mu_N(\omega,u)$ for $\frac{1}{N}=0$ has been used in Eq. (4) to calculate the reduced free energy per monomer.}
\label{1f}
\end{center}
\end{figure}
\begin{figure}[htp!]
\begin{center}
\setlength{\unitlength}{1cm}
\begin{picture}(5,8)
\put(-1,-.1){\scalebox{.8}[.8]{\hspace{-5cm}\includegraphics*{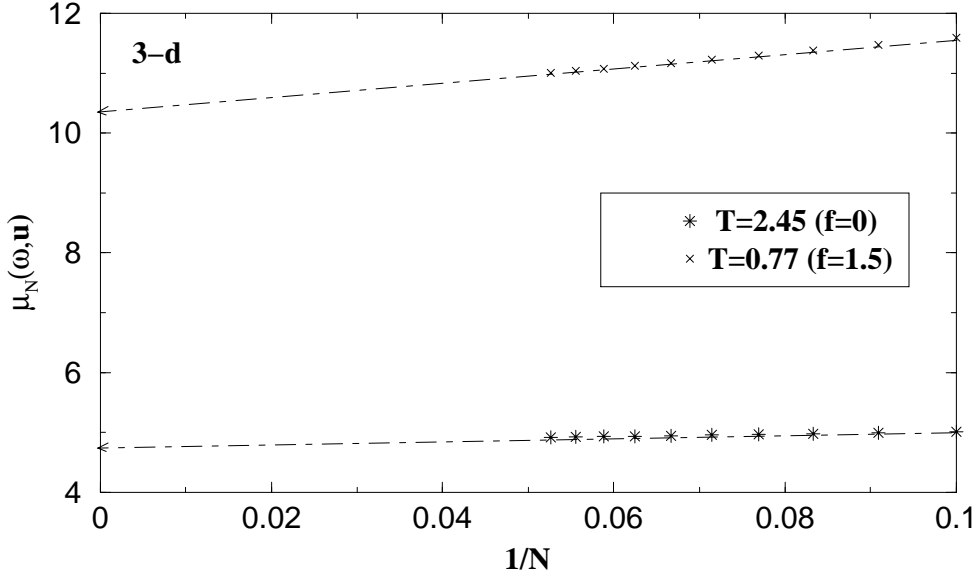}}}
\end{picture}
\caption[a]{ Same as in Fig. (7) but in 3-$d$ and at values of ($f$,$T$)= 
(0,2.45) and (1.5,0.77).}
\label{1f}
\end{center}
\end{figure}

\begin{figure}[htp!]
\begin{center}
\setlength{\unitlength}{1cm}
\begin{picture}(7,8)
\put(-1,-.1){\scalebox{.8}[.8]{\hspace{-5cm}\includegraphics*{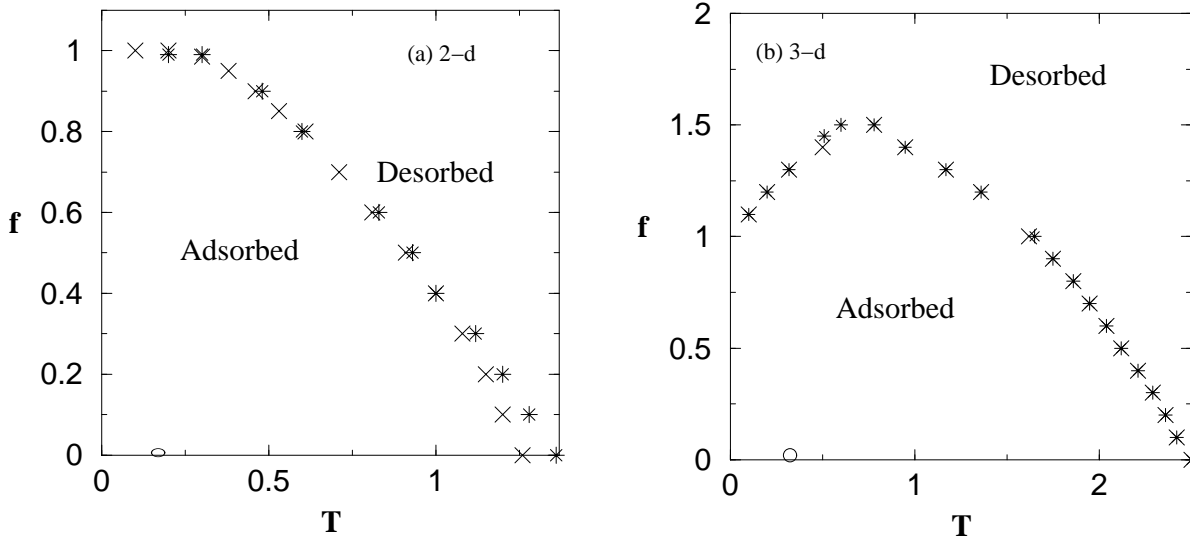}}}
\end{picture}
\caption{The dependence of critical force $f_c(T)$ on $T$ in $(a)$ two and $(b)$ three
-dimensions. The star corresponds to results obtained from extrapolated values of the
reduced free energy 
and cross corresponds to $N=20$ (for 3-$d$) and  $N=31$ (in 2-$d$) respectively.
In 3-$d$ the re-entrance is observed for $f\ge 1$ and in 2-$d$ $f_c(T)$ becomes constant
equal to 1 for $T\le 0.29$.}
\label{1g}
\end{center}
\end{figure}
\newpage
\begin{figure}[htp!]
\begin{center}
\setlength{\unitlength}{1cm}
\begin{picture}(7,9)
\put(-1,-.1){\scalebox{.9}[1.2]{\hspace{-5cm}\includegraphics*{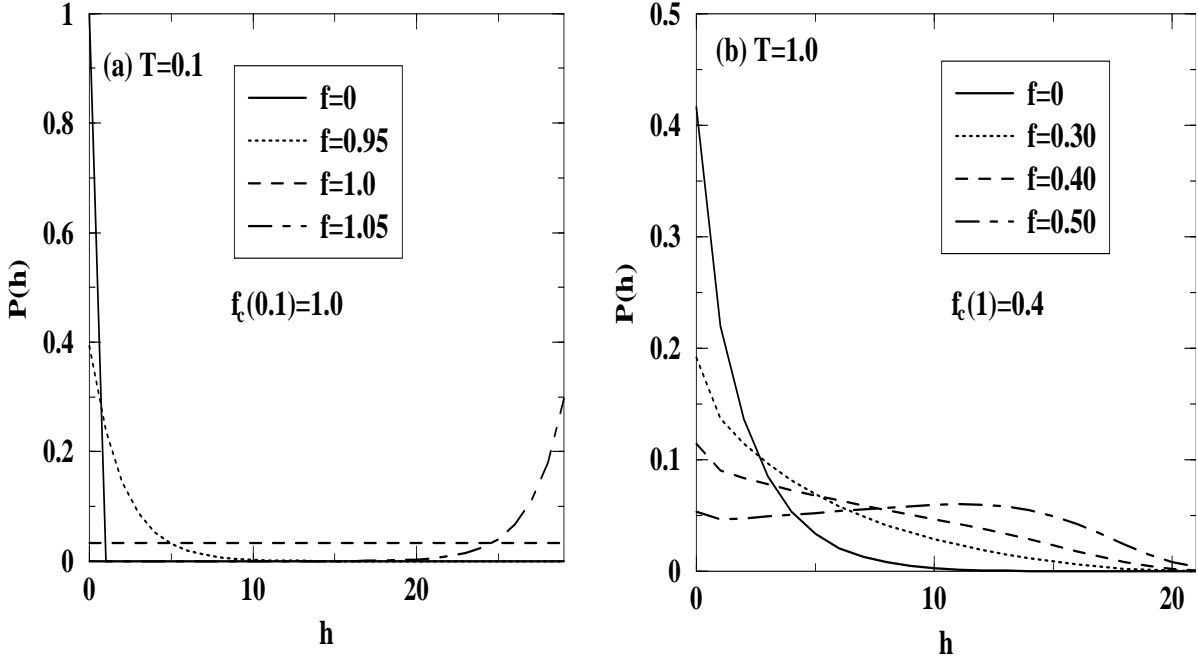}}}
\end{picture}
\caption{The probability distribution curve for the height of the end monomer at which
force is applied from the surface
in 2-$d$ at $T=0.1$ (a) and $T=1.0$ (b). The results given in $(a)$ are for $N=31$ and in
$(b)$ for $N=20$.}
\label{1b}
\end{center}
\end{figure}
\vspace{1.8cm}
\begin{figure}[htp!]
\begin{center}
\setlength{\unitlength}{1cm}
\begin{picture}(7,5)
\put(-1,-.1){\scalebox{.9}[1.0]{\hspace{-5cm}\includegraphics*{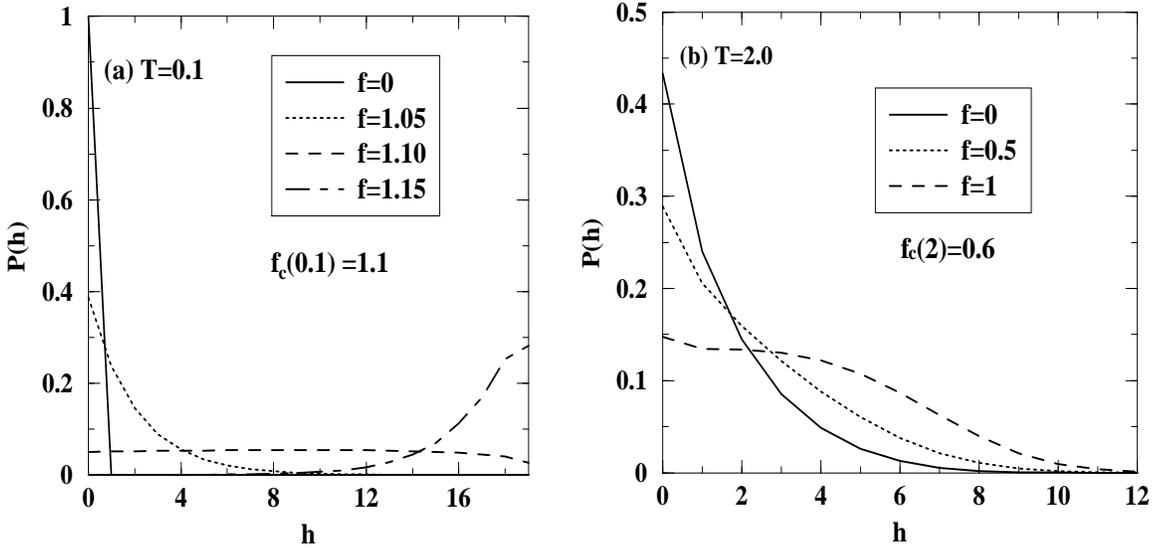}}}
\end{picture}
\caption{The probability distribution curve for the height of the end monomer at which 
force is applied from the surface
in 3-$d$ at $T=0.1$ (a) and $T=2.0$ (b). Other details are same as in Fig. 3.}
\label{1c}
\end{center}
\end{figure}
\begin{figure}[htp!]
\begin{center}
\setlength{\unitlength}{1cm}
\begin{picture}(4,9)
\put(-1,-.1){\scalebox{.9}[.9]{\hspace{-5cm}\includegraphics*{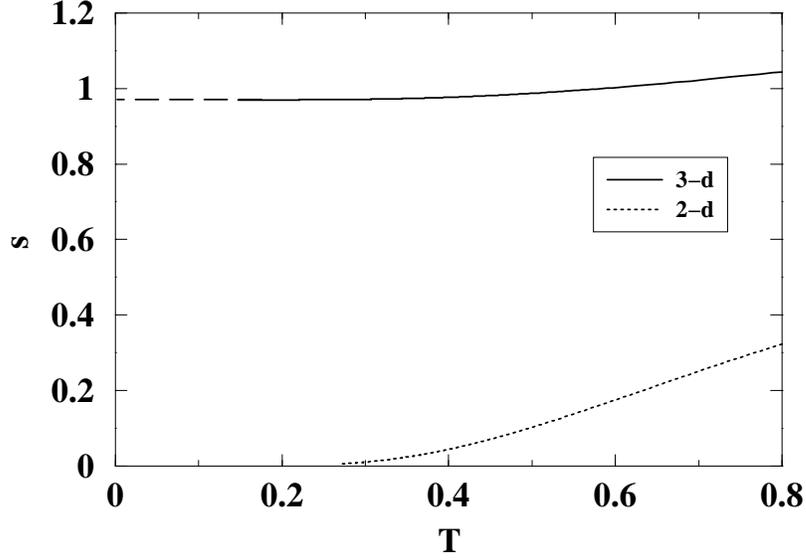}}}
\end{picture}
\caption{The variation of entropy per monomer for $f=0$ in 2-$d$ (dotted line) and in 3-$d$
(full line). The $s$ becomes zero for $T\le 0.27$ in 2-$d$ and becomes nearly constant 
$(\sim 0.97)$ in 3-$d$ for $T\le 0.4$.}
\label{1d}
\end{center}
\end{figure}

\begin{figure}[htp!]
\begin{center}
\setlength{\unitlength}{1cm}
\begin{picture}(6,8)
\put(-1,-.1){\scalebox{.9}[.9]{\hspace{-5cm}\includegraphics*{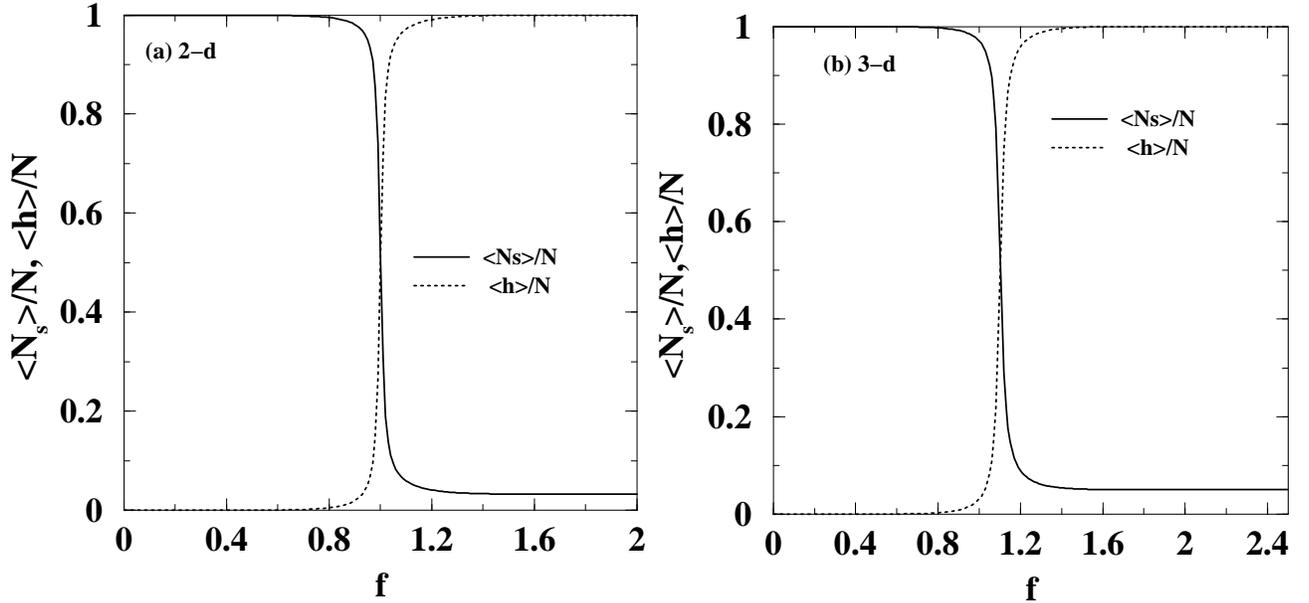}}}
\end{picture}
\caption{The nature of variation of $\frac{<N_s>}{N}$ and $\frac{<h>}{N}$ has been shown 
with $f$ at a fixed value of temperature ($T=0.1$) in $(a)$ 2-$d$ and $(b)$ 3-$d$.
The results given in $(a)$ are for $N=31$ and in $(b)$ $N=20$.}
\label{1e}
\end{center}
\end{figure}

\begin{table}

\vspace{1.5cm}

\caption{ Values of $\mu_N(\omega,u)$}

 in 2-$d$     for $f=0$  and $T=1.3$

 \hspace{2.1cm}  $N$\hspace{3.2cm}  $\mu_N(\omega,u)=\sqrt{\frac{Z_N(\omega,u)}{Z_{N-2}(\omega,u)}}$ \hspace{0.5cm}$\mu_N(\omega,u)={\frac{Z_N(\omega,u)}{Z_{N-1}(\omega,u)}}$

   16     \hspace{3cm}   2.85363  \hspace{3cm}      2.86096     

   17     \hspace{3cm}   2.84831    \hspace{3cm}    2.83571     

   18     \hspace{3cm}   2.84134   \hspace{3cm}     2.84697     

   19     \hspace{3cm}   2.83707   \hspace{3cm}     2.82721     

   20    \hspace{3cm}    2.83164   \hspace{3cm}     2.83607     

   21    \hspace{3cm}    2.82815   \hspace{3cm}     2.82026     

   22     \hspace{3cm}   2.82382   \hspace{3cm}     2.82739     

   23    \hspace{3cm}    2.82093  \hspace{3cm}      2.81449     

   24    \hspace{3cm}    2.81741   \hspace{3cm}     2.82034     

   25    \hspace{3cm}    2.81499   \hspace{3cm}     2.80966     

   26    \hspace{3cm}    2.81208   \hspace{3cm}     2.81451     

   27     \hspace{3cm}   2.81003  \hspace{3cm}      2.80556     

   28     \hspace{3cm}   2.80760 \hspace{3cm}       2.80965     

   29    \hspace{3cm}    2.80585 \hspace{3cm}       2.80206     

   30    \hspace{3cm}    2.80380  \hspace{3cm}      2.80554     

   31     \hspace{3cm}   2.80229  \hspace{3cm}      2.79905     

\end{table}
 	
\begin{table}

\caption{ Values of $\mu_N(\omega,u)$}

 in 2-$d$     for $f=0.8$  and $T=0.6$
 
 \hspace{2.1cm}  $N$\hspace{3.2cm}  $\mu_N(\omega,u)=\sqrt{\frac{Z_N(\omega,u)}{Z_{N-2}(\omega,u)}}$ \hspace{.5cm}$\mu_N(\omega,u)={\frac{Z_N(\omega,u)}{Z_{N-1}(\omega,u)}}$

   16   \hspace{3cm}     5.88172 \hspace{3cm}       5.86876     

   17     \hspace{3cm}   5.85734  \hspace{3cm}      5.84595     

   18     \hspace{3cm}   5.83590  \hspace{3cm}      5.82587     

   19     \hspace{3cm}   5.81692   \hspace{3cm}     5.80798     

   20     \hspace{3cm}   5.79998   \hspace{3cm}     5.79198     

   21     \hspace{3cm}   5.78478   \hspace{3cm}     5.77758     

   22     \hspace{3cm}   5.77106   \hspace{3cm}     5.76454     

   23    \hspace{3cm}    5.75861   \hspace{3cm}     5.75269     

   24     \hspace{3cm}   5.74727   \hspace{3cm}     5.74186     

   25    \hspace{3cm}    5.73690   \hspace{3cm}     5.73194     

   26     \hspace{3cm}   5.72737   \hspace{3cm}     5.72281     

   27     \hspace{3cm}   5.71859   \hspace{3cm}     5.71438     

   28    \hspace{3cm}    5.71047   \hspace{3cm}     5.70657     

   29     \hspace{3cm}   5.70294   \hspace{3cm}     5.69932     

   30     \hspace{3cm}   5.69594  \hspace{3cm}      5.69257     

   31     \hspace{3cm}   5.68942   \hspace{3cm}     5.68627     

\end{table}

\begin{table}

\caption{ Values of $\mu_N(\omega,u)$}

 in 3-$d$     for $f=0$  and $T=2.45$

 \hspace{2.1cm}  $N$\hspace{3.2cm}  $\mu_N(\omega,u)=\sqrt{\frac{Z_N(\omega,u)}{Z_{N-2}(\omega,u)}}$ \hspace{.5cm}$\mu_N(\omega,u)={\frac{Z_N(\omega,u)}{Z_{N-1}(\omega,u)}}$

   10  \hspace{3cm}    5.02935\hspace{3cm}      5.03992

   11    \hspace{3cm}    5.01239  \hspace{3cm}      4.98502     

   12    \hspace{3cm}    4.99209  \hspace{3cm}      4.99918     

   13    \hspace{3cm}    4.97988   \hspace{3cm}     4.96066     

   14    \hspace{3cm}    4.96572   \hspace{3cm}     4.97078     

   15    \hspace{3cm}    4.95643   \hspace{3cm}     4.94212     

   16    \hspace{3cm}    4.94593   \hspace{3cm}     4.94974     

   17     \hspace{3cm}   4.93860   \hspace{3cm}     4.92747     

   18      \hspace{3cm}  4.93047    \hspace{3cm}    4.93346     

   19     \hspace{3cm}   4.92453    \hspace{3cm}    4.91561     

   20     \hspace{3cm}   4.91803    \hspace{3cm}    4.92046     

\end{table}

\begin{table}
\caption{ Values of $\mu_N(\omega,u)$}

 in 3-$d$     for $f=0.77$  and $T=1.5$

 \hspace{1.5cm}  $N$\hspace{3.2cm}  $\mu_N(\omega,u)=\sqrt{\frac{Z_{N}(\omega,u)}{Z_{N-2}(\omega,u)}}$ \hspace{.5cm}$\mu_N(\omega,u)={\frac{Z_N(\omega,u)}{Z_{N-1}(\omega,u)}}$

   10   \hspace{3cm}    11.73428  \hspace{3cm}     11.662856    

   11    \hspace{3cm}    11.59258   \hspace{3cm}     11.52274     

   12    \hspace{3cm}    11.47432    \hspace{3cm}    11.42611     

   13    \hspace{3cm}    11.37927    \hspace{3cm}    11.33262     

   14    \hspace{3cm}    11.29805    \hspace{3cm}    11.26358     

   15     \hspace{3cm}   11.23038    \hspace{3cm}    11.19728     

   16     \hspace{3cm}   11.17161    \hspace{3cm}    11.14599     

   17     \hspace{3cm}   11.12143    \hspace{3cm}    11.09692     

   18     \hspace{3cm}   11.07728    \hspace{3cm}    11.05767     

   19     \hspace{3cm}   11.03893   \hspace{3cm}     11.02022     

   20     \hspace{3cm}   11.00485   \hspace{3cm}     10.98951

\end{table}

\end{document}